# X-type stacking in cross-chain antiferromagnets

Shui-Sen Zhang,[1,2,3] Zi-An Wang,[3,2] Bo Li,[4] Yuan-Yuan Jiang,[3,2] Shu-Hui Zhang,[5] Rui-Chun Xiao,[6] Lan-Xin Liu,[3,2] X. Luo,[3] W. J. Lu,[3] Mingliang Tian,[1,7] Y. P. Sun,[1,3,8] Evgeny Y. Tsymbal,[9,*] Haifeng Du,[1,6,†] and Ding-Fu Shao[3,‡]

[1] *Anhui Province Key Laboratory of Low-Energy Quantum Materials and Devices, High Magnetic Field Laboratory, HFIPS, Chinese Academy of Sciences, Hefei, Anhui 230031, China*

[2] *University of Science and Technology of China, Hefei 230026, China*

[3] *Key Laboratory of Materials Physics, Institute of Solid State Physics, HFIPS, Chinese Academy of Sciences, Hefei 230031, China*

[4] *Institute for Advanced Study, Tsinghua University, Beijing, 100084, China*

[5] *College of Mathematics and Physics, Beijing University of Chemical Technology, Beijing 100029, China*

[6] *Institute of Physical Science and Information Technology, Anhui University, Hefei 230601, China*

[7] *School of Physics and Materials Science, Anhui University, Hefei 230601, China*

[8] *Collaborative Innovation Center of Microstructures, Nanjing University, Nanjing 210093, China*

[9] *Department of Physics and Astronomy & Nebraska Center for Materials and Nanoscience, University of Nebraska, Lincoln, Nebraska 68588-0299, USA*

[*] tsymbal@unl.edu;   [†] duhf@hmfl.ac.cn;   [‡] dfshao@issp.ac.cn

Physical phenomena in condensed matter normally arise from the collective effect of all atoms, while selectively addressing a lone atomic sublattice by external stimulus is elusive. The later functionality may, however, be useful for different applications due to a possible response being different from that occurring when the external stimulus affects the whole solid. Here, we introduce cross-chain antiferromagnets, where the stacking of two magnetic sublattices form a pattern of intersecting atomic chains, supportive to the sublattice selectivity. We dub this antiferromagnetic (AFM) stacking X-type and demonstrate that it reveals unique spin-dependent transport properties not present in conventional magnets. Based on high-throughput analyses and computations, we unveil three prototypes of X-type AFM stacking and identify 15 X-type AFM candidates. Using $\beta$-Fe$_2$PO$_5$ as a representative example, we predict the sublattice-selective spin-polarized transport driven by the X-type AFM stacking, where one magnetic sublattice is conducting, while the other is not. As a result, a spin torque can be exerted solely on a single sublattice, leading to unconventional ultrafast dynamics of the Néel vector capable of deterministic switching of the AFM order parameter. Our work uncovers a previously overlooked type of magnetic moment stacking and reveals sublattice-selective physical properties promising for high-performance spintronic applications.

The alignment of magnetic moments in magnetically ordered materials determines their physical properties. For example, ferromagnets have parallel-aligned magnetic moments producing net magnetization that can be controlled by an applied magnetic field and is responsible for various spin-dependent transport properties. In contrast, antiferromagnets have magnetic orderings with atomic moments compensating each other and resulting in zero net magnetization[1,2]. Due to the obvious difficulties in addressing a single magnetic sublattice by common electric or magnetic means, antiferromagnets seem to be spin-independent, not much promising for applications[3].

Despite this seemingly limiting constraint, a plethora of interesting and potentially useful properties of antiferromagnets has been discovered recently. Among them are spin currents, spin torques, and tunneling magnetoresistance effects[4–18]—key phenomena of spintronics. These advances have been stimulated by symmetry-based analyses which helped, in particular, identify antiferromagnets with momentum-dependent spin polarizations, including spin-split collinear antiferromagnets[19–28], dubbed altermagnets[26,27], and noncollinear antiferromagnets[4,7,14–16].

Although the discovered new properties make antiferromagnets potentially useful[29,30], much more exciting opportunities for new physics and spintronic applications would emerge if one of the magnetic sublattices in antiferromagnets could be selectively addressed by electric means. This yet elusive perspective could not, however, be derived based on the global symmetry of antiferromagnets that relates their macroscopic properties to the spin-dependent electronic structure in momentum space but requires a real space consideration of their magnetic sublattice stacking.

Here, we introduce a new type of collinear antiferromagnetic (AFM) stacking, where two magnetic sublattices form a pattern of intersecting atomic chains. The associated cross-chain antiferromagnets, which we dub X-type antiferromagnets, reveal unique spin-dependent transport properties not known in conventional magnets. Based on high-throughput analyses and computations, we unveil three prototypes of X-type AFM stacking and identify 18 X-type AFM candidates. Using $\beta$-Fe$_2$PO$_5$ as a representative example, we predict that X-type AFM stacking supports sublattice-selective spin-dependent transport, where for a chosen transport direction, one magnetic sublattice is conducting, while the other is not. As a result of this unique property, a spin torque can be exerted solely on a single sublattice, leading to unconventional ultrafast dynamics of the Néel vector capable of deterministic switching of the AFM order parameter. Our results indicate the unprecedented opportunities for fundamental science and practical applications offered by the X-type AFM stacking.

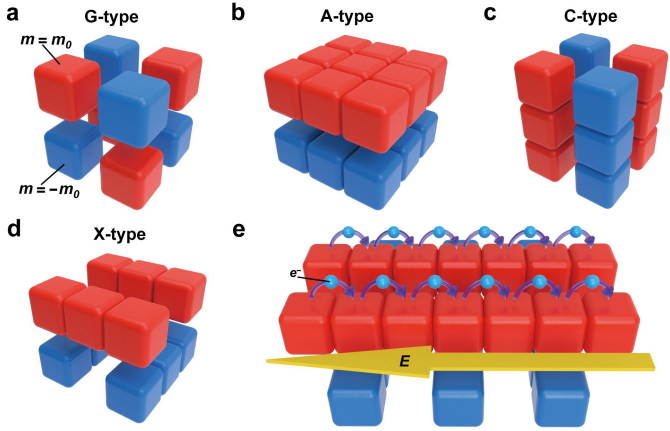

**Fig. 1: "Building blocks" for antiferromagnets.** (**a**) A G-type AFM stacking with the isolated magnetic moments aligned antiparallel between all nearest neighbors. (**b**) An A-type AFM stacking with the antiparallel-aligned FM layers. (**c**) A C-type AFM stacking with the antiparallel-aligned FM chains. (**d**) An X-type AFM stacking with the alternating layers of FM chains parallel to each other within the plane but crossing in the adjacent layers. (**e**) The sublattice selective spin transport allowed by the X-type AFM stacking.

## Results

**Conventional AFM stacking.** Similar to ferromagnets and antiferromagnets being distinguished by their magnetic moment alignments, collinear antiferromagnets on their own are distinguished by different types of AFM stacking. In general, an antiferromagnet is a magnetically-ordered crystal characterized by three attributes: 1) equal magnitudes of magnetization of magnetic sublattices, $|\boldsymbol{m}_\alpha| = |\boldsymbol{m}_\beta|$, where $\alpha$ and $\beta$ are sublattice indices; 2) vanishing net magnetization, $\boldsymbol{M} \propto \sum_\alpha \boldsymbol{m}_\alpha = 0$; and 3) a long-range AFM order, corresponding to a Heisenberg model with energy $H = -\sum_{\alpha \neq \beta} J_{\alpha\beta} \boldsymbol{m}_\alpha \boldsymbol{m}_\beta$ and a negative exchange parameter $J_{\alpha\beta} < 0$ between different sublattices. In collinear antiferromagnets, the two magnetic sublattices can have different structures and stacking patterns.

For example, Fig. 1 (a-c) illustrate typical G-, A-, and C-types of AFM stacking which have been originally proposed for simple cubic systems[31], but now widely used to characterize a broad range of antiferromagnets. For a G-type antiferromagnet, basic "building blocks" are isolated magnetic moments aligned antiparallel between all nearest neighbors (Fig. 1(a)). In this sense, a G-type antiferromagnet can be considered as "genuine," due to AFM coupling solely controlling this type of stacking with ferromagnetic (FM) coupling playing a secondary role. On the contrary, for A-type or C-type antiferromagnets, where the building blocks are antiparallel-aligned two-dimensional (2D) FM-ordered layers (Fig. 1(b)) or one-dimensional (1D) FM-ordered atomic chains (Fig. 1(c)). Such a classification has been used over 70 years[31] and involved other types of conventional AFM stacking such as D-, E-, and F-types, which represent mixtures of parallel and orthogonal FM-aligned chains (D and F) combined with isolated magnetic moments antiparallel to all their nearest neighbors (E).

The classification of AFM stacking is useful to derive sublattice dependent properties. As we discussed in our recent paper[17], some AFM crystals exhibiting a C-type AFM stacking, e.g., those which belong to the rutile family, have small intra-sublattice distances resulting in strong intra-sublattice electron hopping between the neighboring sites. A similar property is expected for certain A-type antiferromagnets. Therefore, from the electronic transport perspective, such A-type and C-type antiferromagnets qualitatively represent FM-ordered sublattices electrically connected in parallel. They support staggered Néel spin currents capable of driving spin-transfer torques and sizable TMR effects if used as electrodes in AFM tunnel junctions[17,18].

**X-type AFM stacking.** Antiferromagnets with stacking patterns other than mentioned above may exhibit even more exciting spin-dependent transport phenomena. As an example, let us consider 1D FM-ordered chains of atoms with opposite magnetic moment orientations (denoted as FM$_A$ and FM$_B$) as the basic building blocks of a new type of AFM stacking. Instead of alternating FM$_A$ and FM$_B$ chains parallel to each other as in the C-type stacking, the new stacking order hosts alternating layers of FM$_A$ and FM$_B$ chains parallel to each other within the plane but crossing in the adjacent layers (Fig. 1(d)). Unlike the A-type AFM stacking with a similar FM ordering within the alternating planes, the new stacking has a sizable separation between the chains making them rather isolated in each plane. This does not violate the three above-mentioned attributes of the AFM order and thus should exist in nature. Due to FM$_A$ and FM$_B$ chain crossings in the antiferromagnets with such stacking, we dub them as *cross-chain* antiferromagnets, or *X-type* antiferromagnets for short.

The proposed X-type AFM stacking is expected to reveal unconventional transport properties never discovered before.

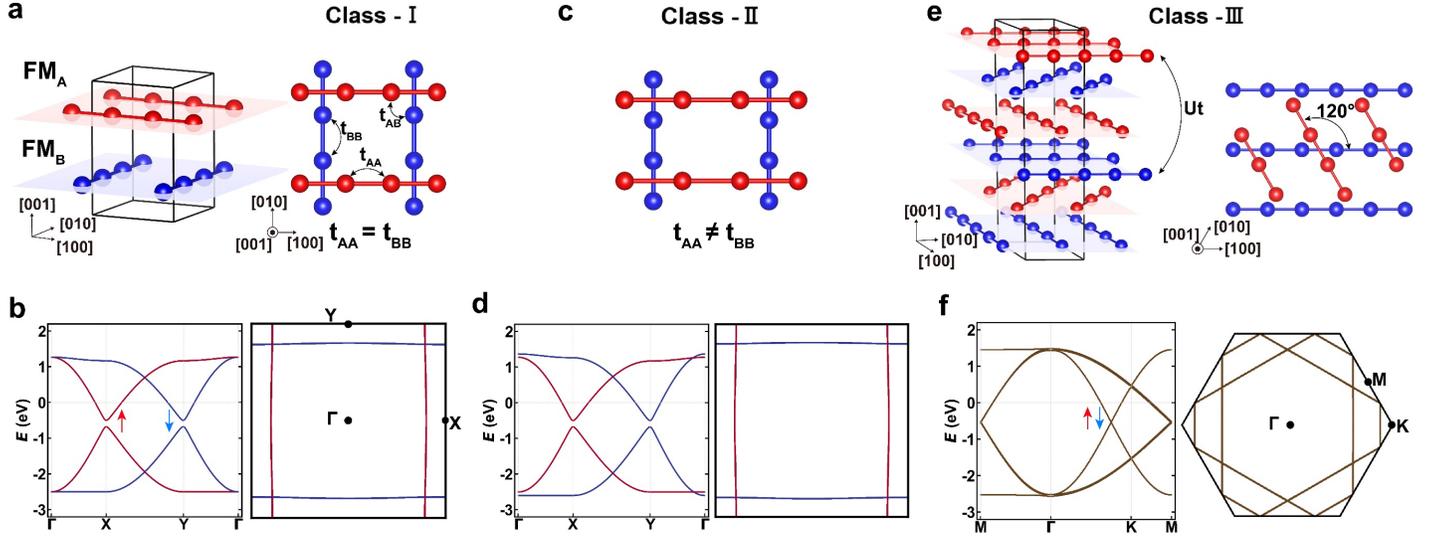

**Fig. 2: Prototypes of X-type AFM Stacking.** (**a**) The side view (left) and the top view (right) of the X-type AFM stacking of class-I, where the magnetic unit cell is the same as the crystal unit cell and the $FM_A$ and $FM_B$ chains are connected by crystal symmetry, leading to equal intra-chain hoppings. (**b**) The band structure with an altermagnetic spin splitting (left) and the associated Fermi surface (right) supported by the class-I X-type AFM stacking with $t_{AA} = t_{BB}$. (**c**) The top view of the X-type AFM stacking of class-II, where the magnetic unit cell is the same as the crystal unit cell, but the $FM_A$ and $FM_B$ chains are not connected by symmetry, leading to different intra-chain hoppings. (**d**) The band structure with a non-altermagnetic spin splitting (left) and the associated Fermi surface (right) supported by the class-II X-type AFM stacking with $t_{BB} = 1.05 t_{AA}$. (**e**) The side view (left) and the top view (right) of the X-type AFM stacking of class-III, where the magnetic unit cell doubles the crystal unit cell resulting in the $\hat{U}\hat{t}$ symmetry combining spin rotation $\hat{U}$ and half-unit-cell translation $\hat{t}$. (**f**) The spin degenerate band structure (left) and the Fermi surface (right) supported by the class-III X-type AFM stacking. In the calculations (b,d,f), we set $t_{AA} = 1$ eV, $t_{AB} = 0.3$ eV, $J = 3$ eV, and $\varepsilon_0 = 2.5$ eV. $t_{AA} = t_{BB}$ is assumed in (f).

Due to the pronounced anisotropy associated with the two cross-chain sublattices, X-type AFM metals are anticipated to exhibit strong anisotropy in their transport properties. For example, when applying an electric field along the $FM_A$ direction, the $FM_A$ chains are to be much more conductive for charge transport than the $FM_B$ chains, provided by the strong intra-chain and weak inter-chain couplings (Fig. 1(e)). In addition, due to the chains being FM-ordered, the current flowing along the $FM_A$ chain must be spin-polarized. Therefore, unlike the A-type and C-type AFM stackings supporting staggered Néel spin currents on both sublattices, the X-type AFM stacking is expected to support isolated Néel spin currents selectively flowing along a single sublattice, provided that the two chains are orthogonal. Similarly, when applying an electric field along the $FM_B$ direction, oppositely spin-polarized Néel spin currents appear in the $FM_B$ chains, while the $FM_A$ chains become non-conducting.

**Prototypes of X-type AFM stacking.** Based on a high-throughput search within the dataset of Material Project[32], we have identified more than 600 crystals composed of alternating layers of chains parallel to each other within the plane but crossing in the adjacent layers. Assuming that the adjacent layers of chains are AFM-coupled in these cross-chain crystals, we identify three prototypes of the X-type AFM stacking order (Fig. 2). The first two prototypes, indicated as class-I and class-II, host the same magnetic unit cell as the crystal unit cell. They can be described by a minimum tight-binding toy model (Fig. 2(a,c)), as follows:

$$H = H_{AA} + H_{BB} + H_{AB}, \quad (1)$$

where $H_{AA}(H_{BB})$ is the Hamiltonian of an isolated FM chain A(B), and $H_{AB}$ is the coupling between FM chains A and B:

$$H_{AA} = t_{AA} \sum_{\alpha,\beta,\langle i,j \rangle} (c^\dagger_{\alpha,i} c_{\beta,j} + h.c.) - J \sum_{\alpha,i} c^\dagger_{\alpha,i} \boldsymbol{\sigma} \cdot \boldsymbol{m}_\alpha c_{\alpha,i}$$
$$+ \varepsilon_0 \sum_{\alpha,i} c^\dagger_{\alpha,i} c_{\alpha,i},$$
$$H_{BB} = t_{BB} \sum_{\gamma,\delta,\langle i,j \rangle} (c^\dagger_{\gamma,i} c_{\delta,j} + h.c.) - J \sum_{\gamma,i} c^\dagger_{\gamma,i} \boldsymbol{\sigma} \cdot \boldsymbol{m}_\gamma c_{\gamma,i}$$
$$+ \varepsilon_0 \sum_{\gamma,i} c^\dagger_{\gamma,i} c_{\gamma,i},$$
$$H_{AB} = t_{AB} \sum_{\alpha,\gamma,\langle i,j \rangle} (c^\dagger_{\alpha,i} c_{\gamma,j} + h.c.). \quad (2)$$

Here $\alpha, \beta$ ($\gamma, \delta$) denote atomic sites within a unit cell of the FM$_A$ (FM$_B$) chain, $i$ ($j$) are cell indices, $t_{AA}$ ($t_{BB}$) is intra-chain hopping parameter within the FM$_A$ (FM$_B$) chain, $t_{AB}$ is the hopping parameter between the FM$_A$ and FM$_B$ chains, $J$ is the exchange parameter, $\varepsilon_0$ is the on-site energy, $\boldsymbol{\sigma}$ is the Pauli matrix vector, and $\boldsymbol{m}_\alpha (\boldsymbol{m}_\gamma)$ is the magnetic moment on the sites $\alpha (\gamma)$.

Within the X-type AFM stacking order of class-I, FM$_A$ and FM$_B$ are connected by a symmetry operation $\hat{O}$ or $\hat{T}\hat{O}$, where $\hat{O}$ is a crystal symmetry such as a mirror/glide plane $\hat{M}_\parallel$ perpendicular to the $x$-$y$ plane, a two-fold rotation/screw axis $\hat{C}_{2\parallel}$ parallel to the $x$-$y$ plane, or a four-fold screw axis $\hat{C}_{4z}$ around the $z$ direction, and $\hat{T}$ is time reversal (Fig. 2(a)). These symmetries require $t_{AA} = t_{BB}$. This prototype of the X-type AFM stacking hosts an altermagnetic spin splitting (Fig. 2(b)).

The X-type AFM stacking order of class-II, on the other hand, does not have symmetry connection between FM$_A$ and FM$_B$, due to different intra-chain/inter-chain distances or relative rotations/distortions of the structural motifs of the chains (Fig. 2(c)). This leads to a slightly different $t_{AA}$ and $t_{BB}$. In this case, the spin degeneracy at the Γ point is removed (Fig. 2(d)), indicating a non-altermagnetic spin splitting, as was discussed recently[33].

We note that the momentum-dependent spin splitting can be understood based on the X-type AFM stacking in real space. For example, for X-type antiferromagnets with orthogonal cross-chains (Fig. 2(a,c)) where the inter-chain hopping $t_{AB}$ is not strong, the in-plane electron hopping is confined within the chain. Therefore, the two spin polarized bands contributed by FM$_A$ (FM$_B$) are highly dispersive with a crossing in the Γ-X (Γ-Y) direction, but weakly dispersive with a large gap between the Γ-Y (Γ-X) direction, resulting in one-dimensional spin-polarized sheets (Fig. 2(b,d)). This band structure indicates that it can be qualitatively considered as the superposition of the electronic structures of the isolated FM$_A$ (FM$_B$) chains.

In addition, there is an X-type AFM stacking order of class-III where the magnetic unit cell doubles the crystal unit cell, leading to the $\hat{U}\hat{t}$ symmetry combining spin rotation $\hat{U}$ and half-unit-cell translation $\hat{t}$ and compensating the magnetization (Fig. 2(e)). This prototype is spin-degenerate in the nonrelativistic limit, as shown in our model Hamiltonian calculation by introducing additional terms in Eq. (1) coming from the isolated chains and their coupling (Fig. 2(f)). The crystal structures supporting the class-III stacking can be found in trigonal and hexagonal systems where the adjacent cross-chains are connected by symmetry $\hat{T}\hat{C}_{3z}$ combining $\hat{T}$ and three-fold screw axis $\hat{C}_{3z}$ (Fig. 2(e)). Similar to class-I and class-II stackings, their band structures feature flat bands and band crossings (Fig. 2(f)). The class-III stacking also exhibits the sublattice/chain selectivity: by applying an electric field perpendicular to two antiparallel chains in the cell, the staggered Néel spin currents can be selectively generated in the chains oblique to the electric field.

By calculating the exchange parameters of cross-chain materials based on 3$d$ transition metals, we identify 15 X-type AFM candidates; among them 13 belong to class-I and 2 belong to class-II, with 6 being experimentally available[34–40] as listed in Table 1 together with their spin space group[41–43]. We point out that more X-type antiferromagnets can be further designed, including these that belong to class-III, by suitable element doping/substitution of cross-chain crystals in material dataset. Below, we consider the class-I X-type AFM candidate $\beta$-Fe$_2$PO$_5$[39,40] as a representative example to demonstrate the sublattice-selective spin transport.

**X-type AFM candidate $\beta$-Fe$_2$PO$_5$.** Collinear antiferromagnet Fe$_2$PO$_5$ has been discovered more than 30 years ago[39,40,44] but remains not widely known. It has two polymorphs, where the $\alpha$-phase, which has an orthogonal unit cell and a C-type AFM stacking below 250K[44], and $\beta$-phase represents an X-type AFM stacking[39,40]. Fig. 3(a) shows the crystal and magnetic structures of $\beta$-Fe$_2$PO$_5$, where the Fe atoms with upward and downward magnetic moments represent two magnetic sublattices denoted as Fe$_A$ and Fe$_B$, respectively. The atomic structure of $\beta$-Fe$_2$PO$_5$ contains in-plane chains of face-sharing FeO$_6$ octahedra along the [100] and [010] directions of the tetragonal cell, which can be transformed to each other by the $\hat{T}\hat{C}_{4z}$ symmetry transformation, indicating a class-I X-type AFM stacking. The P atoms are located between these chains. The Fe ions have a mixed 2+/3+ valence and a small Fe-Fe intra-chain distance, favoring an intra-chain FM coupling via the double-exchange interaction[45]. These intersecting chains share common corners of the FeO₆ octahedra. The resulting Fe-O-Fe bond angle is ~125°, leading to a super-exchange interaction that favors the inter-chain AFM coupling[46]. With a [001] directional Néel vector, *β*-Fe₂PO₅ belongs to the magnetic space group I4₁′/am′d. A monoclinic distortion of *β*-Fe₂PO₅ reported at low temperature[34] does not influence the X-type AFM stacking. The uniqueness of the X-type AFM stacking, metallic ground state, and the Néel temperature $T_N$ = 408 K[39,40] well above room temperature make *β*-Fe₂PO₅ interesting both for fundamental studies and for practical applications.

**Table 1**: Material candidates with X-type AFM stacking: chemical formula, spin space group, Material Project ID (MP-ID)[31], and prototype class.

| Chemical formula | Spin Space group | MP-ID | Prototype |
|---|---|---|---|
| Fe$_2$PO$_5$[33] | C$^{-1}$2/$^{-1}$c$^\infty$m1 | mp-553913 | |
| Co$_2$PO$_5$ | C$^{-1}$2/$^{-1}$c$^\infty$m1 | mp-1272384 | |
| KVAsO$_5$[34] | P$^1$n$^{-1}$a$^{-1}$2$_1$$^\infty$m1 | mp-1202651 | |
| KVPO$_5$[35] | P$^1$n$^{-1}$a$^{-1}$2$_1$$^\infty$m1 | mp-19517 | |
| SrV$_2$(PO$_5$)$_2$[36] | F$^{-1}$d$^1$d$^{-1}$2$^\infty$m1 | mp-559580 | |
| CaNi$_2$(PO$_5$)$_2$ | F$^{-1}$d$^{-1}$d$^1$d$^\infty$m1 | mp-1378551 | Class-I |
| Al(FeO$_2$)$_2$ | I$^{-1}$4$_1$/$^1$a$^\infty$m1 | mp-1248448 | |
| VPbO$_2$ | I$^{-1}$4$_1$/$^1$a$^\infty$m1 | mp-1208173 | |
| Y(FeO$_2$)$_2$ | I$^{-1}$4$_1$/$^1$a$^\infty$m1 | mp-1405822 | |
| Ca(CoO$_2$)$_2$ | I$^{-1}$4$_1$/$^1$a$^1$m$^{-1}$d$^\infty$m1 | mp-1406903 | |
| Co$_2$PO$_5$[37] | I$^{-1}$4$_1$/$^1$a$^1$m$^{-1}$d$^\infty$m1 | mp-1105140 | |
| Fe$_2$PO$_5$[39,40] | I$^{-1}$4$_1$/$^1$a$^1$m$^{-1}$d$^\infty$m1 | mp-18830 | |
| VBiO$_2$ | I$^{-1}$4$_1$/$^1$a$^1$m$^{-1}$d$^\infty$m1 | mp-1207870 | |
| SrV$_2$(PO$_5$)$_2$ | P$^1$1$^\infty$m1 | mp-723593 | Class-II |
| Co$_2$PO$_5$ | I$^1$m$^1$m$^1$d$^\infty$m1 | mp-1276873 | |

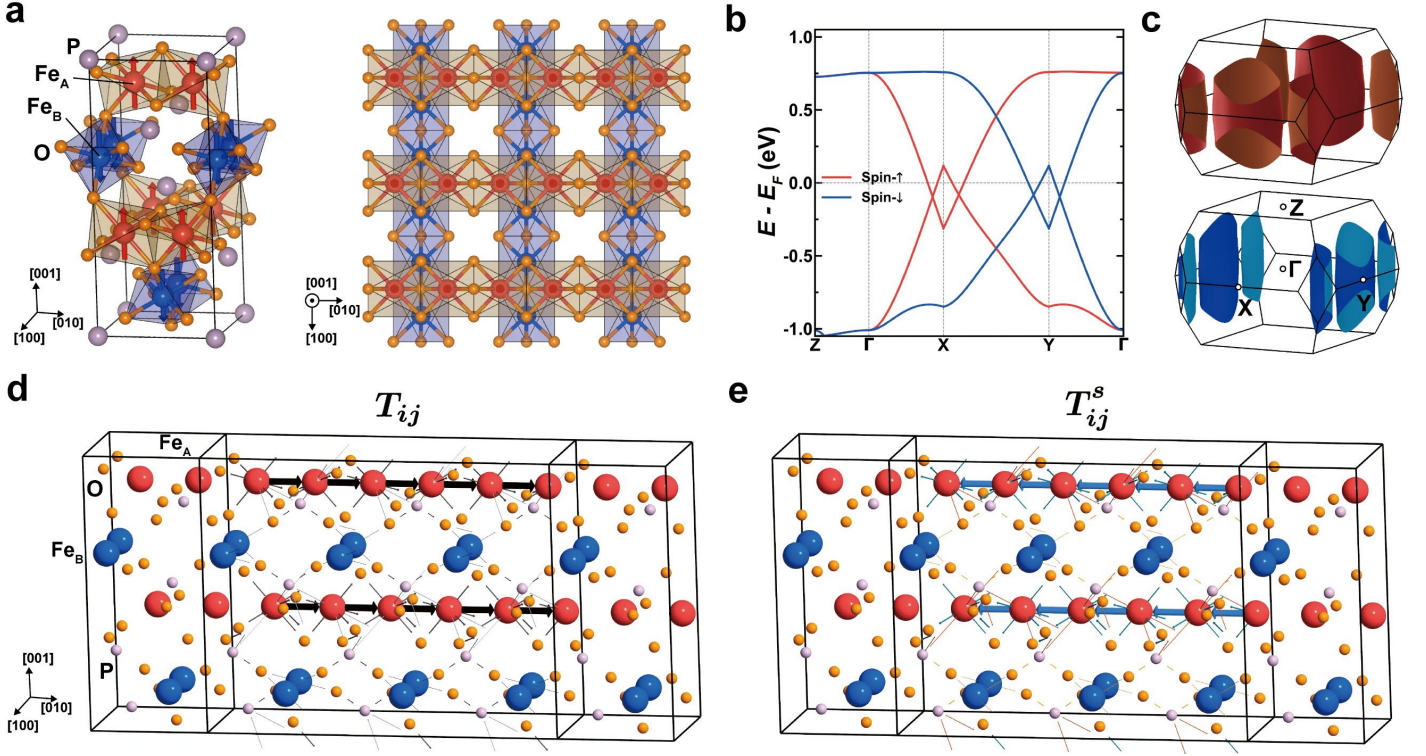

**Fig. 3: X-type antiferromagnet *β*-Fe₂PO₅.** (**a**) *Left*: Crystal and magnetic structures of *β*-Fe₂PO₅. Arrows denote the magnetic moments of Fe atoms. *Right*: Top view of two bottom layers in the unit cell shown in the left. P atoms are ignored. (**b**) Band structure of *β*-Fe₂PO₅. (**c**) The up-spin (top) and down-spin (bottom) Fermi surfaces of *β*-Fe₂PO₅. (**d,e**) Visualization of the calculated bond transmissions $T_{ij}$ (**d**) and bond spin transmissions $T_{ij}^s$ (**e**) of *β*-Fe₂PO₅. The directions of arrows indicate sign of $T_{ij}^{(s)}$ between atoms $i$ and $j$, i.e. a rightward (leftward) arrow imply $T_{ij}^{(s)}$ being positive (negative). The width of the arrows is proportional to $\sqrt{|T_{ij}^{(s)}|}$.

We perform first-principles DFT calculations[47] to explore the electronic and transport properties of *β*-Fe₂PO₅. Fig. 3(b,c) show the calculated nonrelativistic band structure and Fermi surface of *β*-Fe₂PO₅ that reveal giant altermagnetic spin splitting[48] similar to that derived from our toy model (Fig. 2(b)). To elucidate spin-dependent transport properties, we calculate the bond transmission $T_{ij} = T_{ij}^\uparrow + T_{ij}^\downarrow$ and bond spin transmission $T_{ij}^s = T_{ij}^\uparrow - T_{ij}^\downarrow$ for transport along the [010] direction, where ↑ and ↓ denote the two spin channels, and $i$ and $j$ are the atomic indices[49]. As is evident from Fig. 3(d,e), these quantities reflect the anticipated chain-selective transport: both $T_{ij}$ and $T_{ij}^s$ are dominated by the Fe$_A$-Fe$_A$ intra-chain bonds. It is notable that $T_{ij}^s$ are negative, i.e., the spin current is flowing in the opposite direction to the charge current. This is due to the double-exchange interaction between Fe$_A^{2+}$ and Fe$_A^{3+}$ ions, which is mostly mediated by the hopping of electrons with the spins opposite to the Fe$_A$ moments. As a result, the down-spin bands along the Γ-Y direction are highly dispersive (Fig. 3(b)), leading to the down-spin Fermi sheets representative to 1D transmission along the [010] direction (Fig. 3(c)). On the other hand, although there are mobile up-spin electrons on the Fe$_B$ chains to mediate the intra-chain double-exchange interaction, the weak inter-chain coupling due to the large Fe$_A$-Fe$_B$ and Fe$_B$-

Fe$_B$ inter-chain distances leads to negligible $T_{ij}$ and $T_{ij}^S$ associated with Fe$_B$. Therefore, the dispersions of the up-spin bands are weak and have a large energy gap in the Γ-Y direction (Fig. 3(b)). The transport spin polarization in this case is estimated to be as high as ~68% in the ballistic regime, which is further enhanced to ~94% in the diffusive regime, due to a small $T_{ij}^{(s)}$ associated with Fe$_B$ being scattered out[47]. Similarly, for transport along the [100] direction, the total transmission is majorly contributed by the hopping of up-spin electrons between intra-chain Fe$_B$ atoms, corresponding to the band structure along the Γ-X direction and leading the up-spin Fermi surface sheets (Fig. 3(b,c)). Therefore, the spin-split electronic structure of β-Fe$_2$PO$_5$ can be qualitatively considered as the superposition of the electronic structures of the isolated Fe$_A$ and Fe$_B$ chains, as expected based on our toy model.

The strong intra-chain spin transmission supports highly spin-polarized Néel spin currents along the chains in β-Fe$_2$PO$_5$. These Néel spin currents are highly anisotropic due to the X-type AFM stacking. For example, an electric field along the [110] direction produces crossing Néel spin currents with opposite spin polarizations along the Fe$_A$ and Fe$_B$ chains. These currents can be decomposed into a transverse global spin current and the staggered longitudinal Néel spin current hidden in a globally spin-neutral current. On the other hand, when the electric field is along one of the chain directions ([100] or [010]), only the longitudinal chains are conductive producing an isolated Néel spin current, while the transverse chains are almost insulating.

**Spin torques on a single sublattice.** The robust sublattice selective spin-dependent transport property is very interesting, as it opens a possibility to realize the sublattice-resolved spintronic phenomena. As an example, we show that X-type antiferromagnets, such as β-Fe$_2$PO$_5$, allow exerting a spin torque on a single magnetic sublattice. This can be achieved by passing an external spin current along one of the FM-chain directions or by using an isolated Néel spin current on one type of FM chains in the presence of asymmetric boundary conditions to generate self-torque. Assuming that the current is flowing along the FM$_A$ sublattice with magnetic moments $m_A$ pointing to the +z direction, the associated magnetic dynamics of an X-type antiferromagnet can be described by the Landau-Lifshitz-Gilbert-Slonczewski equations[50,51]:

$$\dot{m}_A = \tau_A^F + \tau_A^D + \tau_A^{FL} + \tau_A^{DL},$$
$$\dot{m}_B = \tau_B^F + \tau_B^D. \quad (3)$$

Here $\tau_\alpha^F \propto -m_\alpha \times H_{eff,\alpha}$ and $\tau_\alpha^D \propto m_\alpha \times \dot{m}_\alpha = -m_\alpha \times (m_\alpha \times H_{eff,\alpha})$ are the precession and damping torques on sublattice $\alpha$ induced by the intrinsic effective field $H_{eff,\alpha}$, respectively, and $\tau_A^{FL} \propto -m_A \times H_{st}$ and $\tau_A^{DL} \propto -m_A \times (m_A \times H_{st})$ are the field-like and damping-like torques driven by a spin-torque effective field $H_{st}$ oriented toward the −z direction. Since $H_{eff,\alpha}$ includes the anisotropy field $H_{K,\alpha}$ and the exchange field $H_{E,\alpha}$, the $\tau_\alpha^F$ and $\tau_\alpha^D$ can be further decomposed as $\tau_\alpha^F = \tau_{K,\alpha}^F + \tau_{E,\alpha}^F$ and $\tau_\alpha^D = \tau_{K,\alpha}^D + \tau_{E,\alpha}^D$, where the subscripts $K$ and $E$ denote the associated torques driven by $H_{K,\alpha}$ and $H_{E,\alpha}$, respectively. Due to $H_{E,\alpha}$ being much larger than $H_{K,\alpha}$, $\tau_\alpha^F \approx \tau_{E,\alpha}^F$ and $\tau_\alpha^D \approx \tau_{E,\alpha}^D$ when $m_A$ and $m_B$ are not perfectly antiparallel. Fig. 4(a,b) schematically show how these torques affect the dynamics of $m_A$ and $m_B$. When the isolated spin-torque carrying effective field $H_{st}$ is applied solely to $m_A$, the associated damping-like torque $\tau_A^{DL}$ pulls $m_A$ away from the easy-axis toward the −z direction. This creates canting between $m_A$ and $m_B$, and hence $H_{E,\alpha}$ exerts non-zero torques $\tau_\alpha^F$ and $\tau_\alpha^D$ on both moments $m_A$ and $m_B$. The torque $\tau_\alpha^F$ generates oscillations of the Néel vector, while the torque $\tau_B^D$ pulls $m_B$ toward +z direction even without assistance of an external spin torque. Using realistic parameters $|H_{K,\alpha}| = 2$ T, $|H_{E,\alpha}| = 100$ T, and the damping factor of 0.01, we find that $H_{st}$ much smaller than $H_{K,\alpha}$ ($0.17|H_{K,\alpha}| < |H_{st}| < 0.24|H_{K,\alpha}|$) is sufficient to switch the Néel vector (Fig. 4(c)). The switching trajectory is similar to that for the spin-transfer torque in ferromagnets[50,51], indicating the decisive role of the damping-like torque $\tau_A^{DL}$. Since this $H_{st}$ can be driven by an external spin current, the AFM domains can be aligned according to the spin polarization of the external spin source. This solves the challenging problem of AFM spintronics where measurable time-reversal-odd properties (e.g., TMR in AFM tunnel junctions) are strongly diminished by misaligned domains.

Interestingly, we find that a larger $H_{st}$ ($|H_{st}| > 0.24|H_{K,\alpha}|$) can drive ultrafast oscillations of the Néel vector (Fig. 4(d)). This is because of the much stronger canting of $m_A$ and $m_B$ that produces a large out-of-plane component of $\tau_\alpha^F$ competing with $\tau_A^{DL}$. This property distinguishes the dynamics of X-type antiferromagnets from other antiferromagnets, where full switching and ultrafast oscillations cannot be generated using the same type of spin torques[9,10,52].

**Discussion**

Among the 18 antiferromagnets with the X-type AFM stacking we identified, 12 are metallic and hence are expected to support the properties discussed in this work. For the remaining 6 insulating X-type antiferromagnets, the sublattice selective transport properties associated with magnon excitations are possible[53], offering promising opportunities for spintronics with low-energy dissipation and long coherence length.

In addition to the existing X-type antiferromagnets, the X-type AFM stacking may occur in magnets which are not natural antiferromagnets. For example, one can design *X-type ferrimagnets*, with AFM stacking of cross-chains composed of different magnetic elements. In addition, it is possible to engineer *synthetic* X-type antiferromagnets with the same advantages as their natural counterparts.

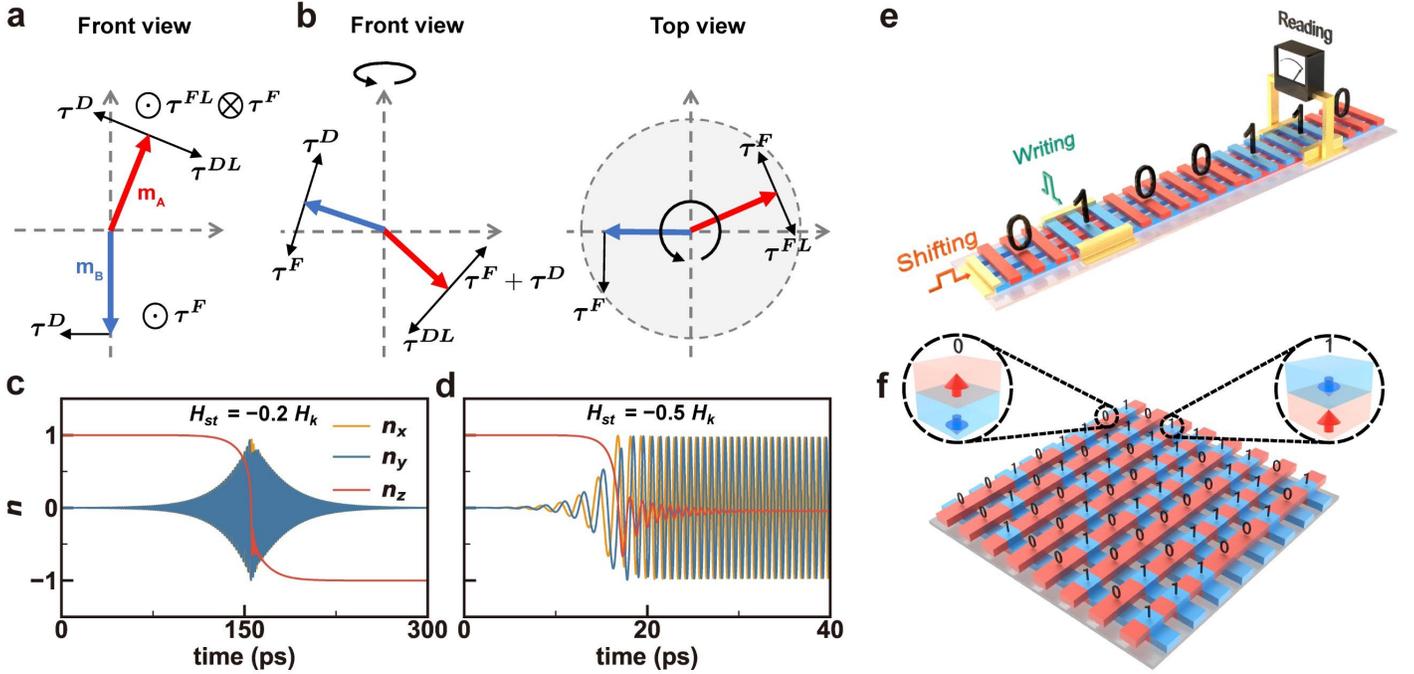

**Fig. 4: Sublattice selective spin torques and their implications for spintronics**. (**a**) Vectors of magnetic moments and spin torques in the initial process of spin dynamics. (**b**) Vectors of magnetic moments and spin torques in the process of Néel vector oscillations driven by the single-sublattice spin torque: front (left) and top (right) views. (**c,d**) Simulated dynamics of the $x$, $y$, and $z$ components of the normalized Néel vector **n** under the current-induced spin-torque effective field $H_{st} = -0.2 H_K$ (**c**) and $H_{st} = -0.5 H_K$ (**d**). (**e**) Schematic of an X-type racetrack memory. Here, the colored bars denote $FM_A$ and $FM_B$ chains of an X-type antiferromagnet, where the upward (downward) magnetic moments are denoted by red (blue) color. In this schematic, we assume with a slight exaggeration that the electrodes contact three chains. (**f**) Schematic of the X-type AFM stacking with crossing atomic chains forming an intrinsic crossbar array. Each bar is composed of a bunch of parallel chains, and the red and blue colors denote upward and downward moment orientations in these chains. The gaps between these bars are the chains not connected to the electrodes. The AFM domains at the intersections of these bars represent "0" and "1" bits.

From the technological point of view, the X-type AFM stacking offers a new approach for non-volatile memories. For example, in an X-type antiferromagnet, one can selectively nucleate and move domains via spin torques by passing an isolated Néel spin current or an external spin current along a chosen $FM_A$ or $FM_B$ chain (Fig. 4(e)). This promises an ultrafast X-type racetrack memory, where the oppositely magnetized AFM domains can serve as "0" and "1" bits of information (Fig. 4(e)). Moreover, the X-type AFM stacking with crossing chains of magnetic atoms forms an intrinsic crossbar array that is widely used in magnetic random-access memories (MRAMs) (Fig. 4(f)). Assuming a bilayer array of $\beta$-Fe$_2$PO$_5$ with one bit area containing a domain of $50 \times 50$ intersecting chains, the associated storage density is ~15 Tb/in$^2$, promising for future big data applications.

Overall, the X-type AFM stacking signifies an exciting field of research for fundamental science and an attractive new paradigm for AFM spintronics. The predicted sublattice-selective electronic transport carrying isolated Néel spin currents has multiple important implications, such as spin torques exerted on a single magnetic sublattice, unconventional ultrafast spin dynamics, and deterministic switching of the AFM domains. These unprecedented functionalities promise new AFM spintronic devices which can be employed in non-volatile memories with ultra-high density and ultra-fast switching speed. We hope therefore that our theoretical results will stimulate the experimentalists working in this field to explore the properties of X-type AFM stacking relevant to spintronics and beyond.

**Acknowledgments**
This work was supported by the National Key R&D Program of China (Grant No. 2021YFA1600200), the National Natural Science Funds for Distinguished Young Scholar (Grant No. 52325105), the National Natural Science Foundation of China (Grants Nos. 12274411, 12241405, and 52250418), the Basic Research Program of the Chinese Academy of Sciences Based on Major Scientific Infrastructures (Grant No. JZHKYPT-2021-08), and the CAS Project for Young Scientists in Basic Research (Grant No. YSBR-084). E.Y.T acknowledges support from the Division of Materials Research of the National Science Foundation (NSF grant No. DMR-2316665). Computations were performed at Hefei Advanced Computing Center.